\documentclass[aps, superscriptaddress, twocolumn, pre, longbibliography]{revtex4-1}
\usepackage{amsmath,graphicx,color,epsfig,latexsym,bm,ulem,dutchcal,subfigure,float,tikz,eucal, mathpazo,times, braket}
\usepackage[colorlinks=true, linkcolor = blue, urlcolor  = blue, citecolor = blue, anchorcolor = red]{hyperref}   
\newcommand{\abs}[1]{\left\vert#1\right\vert}

\usetikzlibrary{positioning}

\begin{document}
	
	
	\title{Engineering entanglement between resonators by hot environment}
	\author{M. Tahir Naseem} 
	\affiliation{Department of Physics, Ko\c{c} University, 34450 Sariyer, Istanbul, T\"urkiye}
	\author{\"Ozg\"ur E. M\"ustecapl\ifmmode \imath \else \i \fi{}o\ifmmode \breve{g}\else \u{g}\fi{}lu}
	\email{omustecap@ku.edu.tr}
	\affiliation{Department of Physics, Ko\c{c} University, 34450 Sariyer, Istanbul, T\"urkiye}
	\affiliation{T\"UB\.{I}TAK  Research  Institute  for  Fundamental  Sciences,  41470  Gebze,  T\"urkiye}

	\begin{abstract}
Autonomous quantum thermal machines do not require an external coherent drive or work input to perform the desired tasks, making them a promising candidate for thermal management in quantum systems. Here, we propose an autonomous quantum thermal machine in which two uncoupled macroscopic mechanical resonators or microwave resonators achieve considerable entanglement via a hot thermal bath. This becomes possible by coupling the resonators to a common two-level system or third harmonic oscillator and driving it by the hot incoherent thermal bath. The critical step to make the entanglement involves suitable engineering of the hot bath, realized by bath spectrum filtering. Our results suggest that the bath spectrum filtering can be an alternative to typical non-autonomous reservoir engineering schemes to create exotic quantum states.
	\end{abstract}
	
	\maketitle

	\section{Introduction}\label{sec:model}
Entanglement plays a key role in the 
 quantum information processing and quantum technologies~\cite{Ac_n_2018}. Therefore, the generation and protection of entanglement in quantum systems is of paramount importance. However, due to the unavoidable coupling of the quantum systems with their environments, entanglement generally degrades or lost. To counter the decoherence caused by system-environment coupling, several schemes have been proposed that exploit the non-unitary evolution of quantum systems to create steady-state entanglement~\cite{PhysRevA.76.062307, PhysRevA.78.042307, Diehl2008, Verstraete2009}. The so-called “reservoir engineering” scheme has been proposed in several physical systems~\cite{PhysRevLett.98.240401, PhysRevA.76.062307, PhysRevA.78.042307, Diehl2008, Verstraete2009, PhysRevA.81.043802, PhysRevE.82.021921, PhysRevLett.106.090502, Arenz_2013, PhysRevB.88.035441, PhysRevA.88.032317, PhysRevLett.111.246802} and experimentally realized in atomic systems~\cite{PhysRevLett.107.080503, Barreiro2011, Lin2013, Shankar2013}. The idea of reservoir engineering is extended to optomechanical systems for the generation of entanglement between optical or mechanical resonators~\cite{PhysRevLett.110.253601, PhysRevA.89.063805, PhysRevA.89.014302, Li_2015}. In reservoir engineering, desired dissipative dynamics is obtained by engineering effective reservoirs using external classical controls or work inputs.

In a different approach, entanglement can be created at thermal equilibrium by coupling a composite quantum system to a cold bath, and sufficient cooling the system may lead to an entangled state~\cite{PhysRevLett.87.017901, PhysRevA.64.042302}. In the absence of thermal equilibrium, the temperature gradient can be used to create entanglement~\cite{PhysRevA.59.2468, PhysRevLett.88.197901, PhysRevA.65.042107, PhysRevA.65.040101, PhysRevLett.89.277901, PhysRevLett.91.070402, PhysRevA.74.052304, PhysRevA.75.032308, PhysRevA.78.062301, PhysRevA.84.012319, Bellomo_2013, Bohr_Brask_2015, PhysRevLett.120.063604, Hovhannisyan_2019, PhysRevA.99.042320, Khandelwal_2020, PhysRevA.101.012315, PhysRevA.104.052426, Zhang2021, PRXQuantum.2.040346}. These works show that the energy current between the reservoirs generates entanglement between the quantum systems~\cite{PhysRevA.101.012315, Bellomo_2013, Bohr_Brask_2015, PhysRevLett.120.063604, Hovhannisyan_2019, Khandelwal_2020}. More specifically, a recent study shows that a critical value of heat flow is required for the emergence of entanglement between two coupled qubits~\cite{Khandelwal_2020}. However, in all these proposals, additional filtering operation or feedback is required to make the amount of entanglement significantly large~\cite{Bohr_Brask_2015, Khandelwal_2020}, with  the exception of a coupled qubits system~\cite{BohrBrask2022operational}. Alternatively, a hot reservoir with effective negative temperature can be employed for entanglement enhancement~\cite{PhysRevLett.120.063604}. These entanglement generation proposals are for composite two-level systems with some exceptions of works based on optical and atomic oscillators~\cite{PhysRevLett.88.197901, Laha:22}, and $d$-dimensional composite systems~\cite{Tavakoli2018heraldedgeneration}. Here, we propose creating entanglement between non-interacting macroscopic mechanical resonators via an effective common hot bath. The higher temperature of the hot bath can create significantly large entanglement without the use of additional feedback or filtering operations.

In this work, we ask these questions: Is it possible to devise a scheme in which dissipative dynamics can be engineered for transient or steady-state entanglement in an autonomous setting? If yes, is it possible to create significant entanglement between uncoupled macroscopic mechanical resonators without the need for additional feedback?
 We answer these questions in the affirmative by employing simple spectral filtering of the thermal baths present in the system. Hence incorporating the advantages of reservoir engineering in autonomous entanglement generation thermal machines. For the implementation of our scheme, we consider two uncoupled mechanical or microwave resonators interacting with 
 a common ancilla system via {\it energy-field} coupling~\cite{RevModPhys.85.623,RevModPhys.86.1391, PhysRevLett.120.227702}. By employing simple reservoir engineering based on spectral filtering of the thermal baths, mechanical resonators dissipatively evolve to an entangled steady-state. 
 Note that spectral filtering to engineer reservoirs differs from typical reservoir engineering schemes~\cite{PhysRevLett.98.240401}, and local filtering operations which are used to enhance entanglment in the out-of-equilibrium systems~\cite{Tavakoli2018heraldedgeneration}. 
 In a typical reservoir engineering scheme, external control or work input is required to obtain desired effective dissipation of the system. Contrary, bath spectrum filtering do not require time-dependent interactions or external classical control, making our entanglement machine truly `quantum.' Bath spectrum filtering was pioneered by G. Kurizki and co-workers~\cite{doi:10.1080/09500349414550381} and applied to enhance several thermal tasks such as output power~\cite{Ghosh12156}, efficiency at maximum power~\cite{Naseem_2020}, thermal management~\cite{PhysRevResearch.2.033285}, simultaneous cooling of resonators~\cite{Naseem2021}, and antibunching of a phonon field~\cite{PhysRevA.105.012201}.

Compared with previous proposals of autonomous thermal entanglement machines, our scheme has several distinct advantages: (i) We use simple filtering of the thermal baths to engineer an effective reservoir that governs the resonators’ dissipative dynamics. It is well-known that the two-mode squeezing interaction with sufficiently low temperature is sufficient in a typical entanglement generation scheme, such as in multimode optomechanical systems~\cite{Li_2015}. However,
our proposal differs entirely from previously reported optomechanics schemes~\cite{PhysRevLett.110.253601, PhysRevA.89.063805, Li_2015}, which rely on external control drives to obtain effective two-mode squeezing interaction between the uncoupled resonators. On the contrary, we employ only incoherent thermal baths to induce two-mode squeezing-like environment-mediated interactions between the resonators.
 (ii) The amount of entanglement increases monotonically with the temperature gradient, allowing entanglement between the resonators in the presence of a hot environment.
 In the previous proposals, there exists an optimal value of temperature gradient for
maximum entanglement, after which entanglement starts decreasing~\cite{PhysRevLett.88.197901, Bohr_Brask_2015}. We note that in some fermionic systems, entanglement increases monotonically with the temperature gradient~\cite{BohrBrask2022operational}. 
(iii) Our machine can create significant entanglement between uncoupled macroscopic mechanical resonators without the need for additional feedback or filtering operation~\cite{Bohr_Brask_2015}. 
(iv) Cooling and entanglement generation between the resonators occur simultaneously in our scheme (see appendix~\ref{App:A} for details).

The rest of the paper is organized as follows. In Sec.~\ref{sec:MS} we describe the model system for the implementation of our scheme. Then in Sec.~\ref{sec:ME} derivation of the master equation is given for the system analysis. The results for the entanglement quantification is discussed in Sec.~\ref{sec:Results}. Finally, we present conclusions of this work in Sec.~\ref{sec:conc}.
 
\section{Model system}\label{sec:MS}
\begin{figure}[t!]
  \centering
  \includegraphics[scale=0.32]{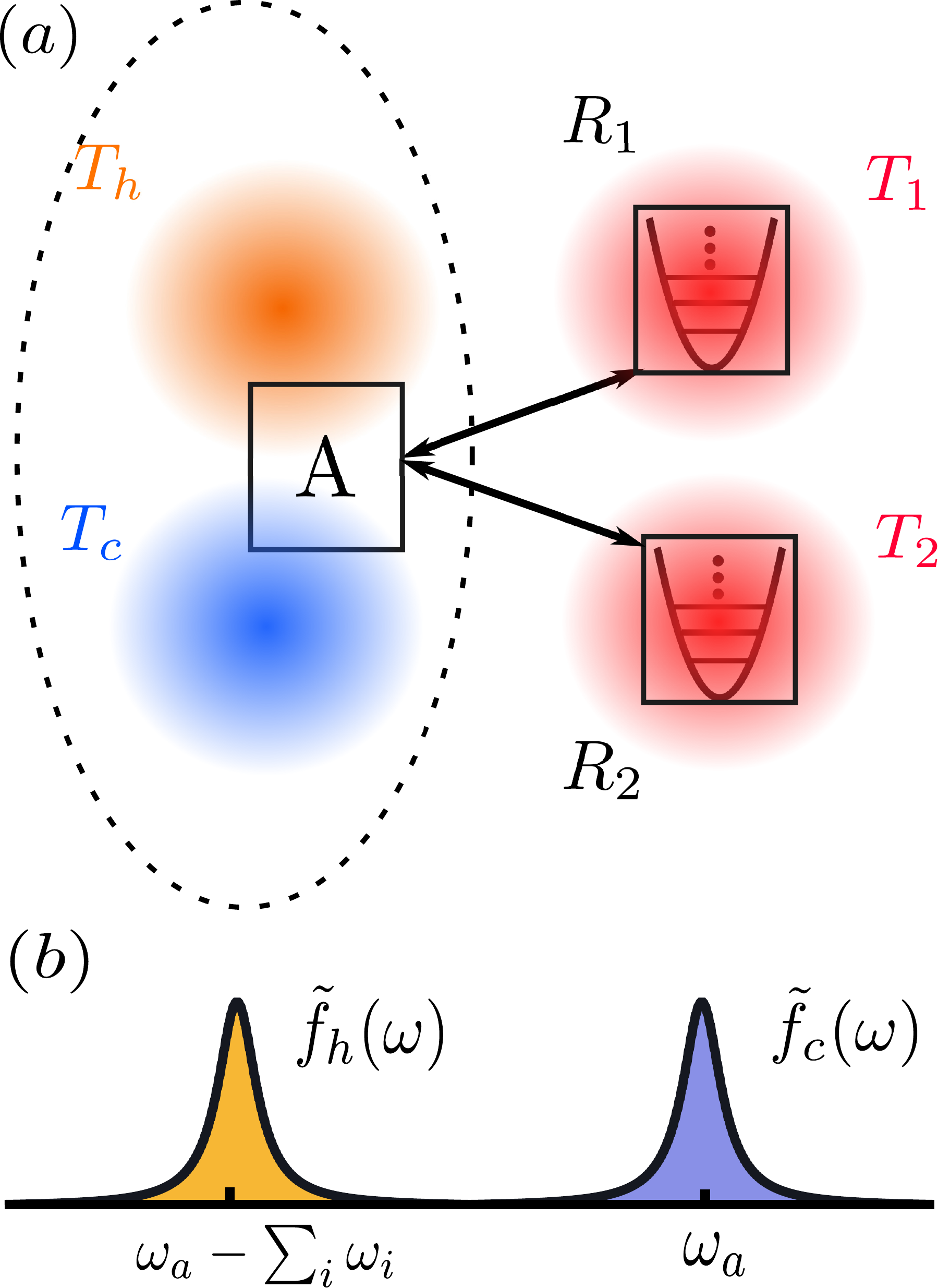}
  \caption{{\bf{(a)}} Model description. Our autonomous entanglement generation thermal machine consists of a two-level system (TLS) or a harmonic oscillator $A$ simultaneously coupled with two harmonic oscillators $R_{i}$ via {\it{energy-field}} interaction~\cite{RevModPhys.86.1391, PhysRevLett.120.227702}. Subsystem $A$ can be a quantum dot~\cite{PhysRevLett.92.075507}, a superconducting qubit~\cite{PhysRevLett.95.097204, PhysRevB.80.144508}, an electronic spin~\cite{PhysRevB.79.041302}, an optical or microwave resonator~\cite{RevModPhys.86.1391}. The harmonic oscillators $R_{i}$ can be microwave~\cite{Bothner2021} or nanomechanical resonators~\cite{LaHaye2009}. The resonators $R_{i}$ are unavoidably coupled to their thermal environments of temperature $T_{i}$, and subsystem $A$ is coupled to two thermal baths of temperatures $T_{h}$ and $T_{c}$. All these thermal baths are independent and can have any non-negative temperature in the limit $T_{h}>T_{i}>T_{c}$. The black dashed oval shows that taking a trace over subsystem $A$ gives an effective joint bath for $R_{i}$, which may creates quantum correlations between the resonators $R_{i}$. {\bf{(b)}} Spectrally filtered baths spectra. $\tilde{f}_{h}$ and $\tilde{f}_{c}$ are the filtered spectra of the hot and cold baths, respectively. The width and center of the filtered bath spectrum can be controlled by the system-bath interaction strength and filter frequency [see Eq.~(\ref{eq:filter})]. The subsystem $A$ and resonators $R_{i}$ frequencies are given by $\omega_{a}$ and $\omega_{i}$, respectively. These spectrally separated filtered baths lead to a dissipative dynamics given in Eq.~(\ref{eq:filtMEM}).}
  \label{fig:fig1}
\end{figure}

In our scheme, we consider a model shown in Fig.~\ref{fig:fig1}. It consists of a two-level system or a harmonic oscillator interacting with the two resonators $R_{1}$ and $R_{2}$ via {\it{longitudinal}} (optomechanical-like) coupling ~\cite{RevModPhys.86.1391}. The Hamiltonian of this system is given by ~\cite{PhysRevLett.107.063601, Massel2012, PhysRevLett.115.203601, PhysRevLett.120.227702, Bothner2021} (we take Plank's constant $\hbar=1$)
\begin{equation}\label{eq:sysHam}
\hat{H}_{S} = \omega_{a}\hat{n}_{a}+\sum_{i=1,2}\omega_{i}\hat{b}_{i}^{\dagger}\hat{b}_{i}+\sum_{i=1,2}g_{i}\hat{n}_{a}(\hat{b}_{i}+\hat{b}_{i}^{\dagger}),
\end{equation}
with $\omega_{a}$ ($\omega_{i}$) being the frequency of the subsystem $A$ ($R_{i}$), and $g_{i}$ is the interaction strength between $A$ and resonators $R_{i}$. The annihilation (creation) operator of the resonator $R_{i}$ is denoted by $\hat{b}_{i}$ ($\hat{b}_{i}^{\dagger}$). $\hat{n}_{a}=\hat{a}^{\dagger}\hat{a}$ is the number operator of the resonator $A$, with $\hat{a}_{i}$ and $\hat{a}_{i}^{\dagger}$ being annihilation and creation operators, respectively. In case of a TLS, $\hat{n}_{a}$ is replaced by Pauli spin matrix $\hat{\sigma}_{z}$ and $\omega_{a}$ is multiplied by a factor of $1/2$ in Eq.~(\ref{eq:sysHam}).

Note that the Hamiltonian given in Eq.~(\ref{eq:sysHam}) can be realized in various setups. For example, the TLS can be an electronic spin qubit~\cite{PhysRevB.79.041302}, a superconducting qubit~\cite{PhysRevLett.95.097204, PhysRevB.80.144508}, a quantum dot~\cite{PhysRevLett.92.075507}, or lower two levels of an optical resonator under weak excitation approximation~\cite{Moqadam2015}. The subsystem $A$ can also be an optical or microwave resonator~\cite{RevModPhys.86.1391, PhysRevLett.120.227702, Bothner2021}.
Further, $R_{i}$ can be nanomechanical resonators~\cite{RevModPhys.86.1391}, or superconducting transmission line resonators~\cite{PhysRevLett.120.227702, Bothner2021}. 
Hence our scheme is not limited to a specific setup. In our numerical results, we use the parameters representative of an electro-mechanical system~\cite{LaHaye2009}. 

In addition to interaction with resonators, subsystem $A$ is coupled with two thermal baths of temperatures $T_{h}$ and $T_{c}$. Each resonator $R_{i}$ is unavoidably coupled to its environment modeled by an ensemble of oscillators with a thermal distribution of temperature $T_{i}$. The need to couple $A$ with two thermal baths is because we shall simultaneously cool the resonators $R_{i}$ to their ground-state before the entanglement generation step. The cooling requires a hot bath which helps to remove energy from resonators and dump it into a cold bath~\cite{Naseem_2020, Naseem2021}. Moreover, our scheme involves bath spectrum filtering to obtain non-overlapping baths spectra [Fig.~\ref{fig:fig1}(b)], in which each bath can only induce transitions of a single frequency. In this case, the energy transfer is only possible if $A$ is coupled with at least two baths~\cite{PhysRevLett.109.090601, Naseem_2020, Naseem2021, PhysRevA.105.012201}. We assume all the baths can attain any non-negative finite temperature provided $T_{h}>T_{i}>T_{c}$. The free Hamiltonian of the thermal baths is given by 
\begin{equation}
\hat{H}_{E} = \sum_{k, E} \omega_{k, E}\hat{c}_{k, E}^{\dagger}\hat{c}_{k, E},
\end{equation}
with $E=h,c, i$, and $\omega_{k,E}$ being the frequency of $k$th mode and $\hat{c}_{k, E}$ ($\hat{c}_{k, E}^{\dagger}$) is corresponding bosonic annihilation (creation) operator. The sum is taken over the infinite number of bath modes indexed by $k$. The interaction of the isolated $A-R_{i}$ system with the environments is given by
\begin{align}\label{eq:Hse}
\hat{H}_{S-E} &= \sum_{k, q=h,c} g_{k, q}(\hat{a} + \hat{a}^{\dagger})(\hat{c}_{k, q}+\hat{c}^{\dagger}_{k,q}) \\ \nonumber   &\quad\qquad + \sum_{k, i} g_{k, i} (\hat{b}_{i}+\hat{b}_{i}^{\dagger})(\hat{c}_{k, i}+\hat{c}^{\dagger}_{k,i}),
\end{align}
with $g_{k,q}$ ($g_{k,i}$) being the coupling strength between the $A$ (resonators $R_{i}$) with the $k$th mode of the bath. In case of a TLS, operators $\hat{a}$  and $\hat{a}^{\dagger}$ are replaced by respective $\hat{\sigma}_{-}$ and $\hat{\sigma}_{+}$ matrices.

\section{The master equation}\label{sec:ME}
The microscopic derivation of the master equation requires first to diagonalize the Hamiltonian in Eq.~(\ref{eq:sysHam}) using the transformation~\cite{1963JETP...16.1301L, PhysRevB.93.134501, PhysRevA.98.052123}
\begin{equation}
\hat{u} = e^{-\hat{n}_{a}\sum_{i}\alpha_{i}(\hat{b}_{i}^{\dagger}-\hat{b}_{i})},
\end{equation}
where $\alpha_{i}=g_{i/\omega_{i}}$. The diagonalized Hamiltonian has the form
\begin{equation}\label{eq:TransOper}
\tilde{H}_{S}=\hat{u}\hat{H}_{S}\hat{u}^{\dagger}= \omega_{a}\tilde{n}_{a}+\sum_{i}\tilde{b}^{\dagger}_{i}\tilde{b}_{i}-\sum_{i}\frac{g^2_{i}}{\alpha_{i}}\tilde{n}^{2}_{a}.
\end{equation}
The transformed operators are given by
\begin{align}\label{eq:transform}
\tilde{a} &= \hat{a} e^{-\sum_{i}\alpha_{i}(\hat{b}_{i}^{\dagger}-\hat{b}_{i})}, \\
\tilde{b}_{i} &= \hat{b}_{i} - \alpha_{i}\hat{n}_{a}.
\end{align}
The master equation can be derived by first transforming the isolated system operators in Eq.~(\ref{eq:Hse}) by using Eq.~(\ref{eq:transform}); then transforming the system operators in Eq.~(\ref{eq:TransOper}) into the interaction picture, and followed by the standard Born-Markov and Secular approximations. The resulting equation has the form~\cite{PhysRevE.90.022102, Naseem2021, PhysRevA.105.012201}
\begin{equation}
\frac{d\tilde{\rho}}{dt}= \tilde{\mathcal{L}}_{q}\tilde{\rho} + \sum_{i}\tilde{\mathcal{L}}_{i}\tilde{\rho}.
\end{equation}
Here the Liouville super-operators for the subsystem $A$ and baths of the resonators $R_{i}$ are denoted by $\tilde{\mathcal{L}}_{q}$ and $\tilde{\mathcal{L}}_{i}$, respectively. These are given by
\begin{widetext}
\begin{eqnarray}
	\label{eq:L_L}
\tilde{ \mathcal{L}}_{q}\tilde{\rho} &=&  f_{q}(\omega_{a})\{D[\tilde{a}]+\alpha^{3}\sum_{i}D[\tilde{a}\tilde{b}_{i}^{\dagger}\tilde{b}_{i}]\}
	 + f_{q}(-\omega_{a})\{D[\tilde{a}^{\dagger}]  + \sum_{i}\alpha^{3}D[\tilde{a}^{\dagger}\tilde{b}_{i}^{\dagger}\tilde{b}_{i}]\}
	 + \sum_{i,n=1,2}\alpha^{n+1}\big\{f_{q}(\omega_{a}-n\omega_{i})D[\tilde{a}\tilde{b}_{i}^{\dagger n}]  
\nonumber\\&&+  f_{q}(-\omega_{a}+n\omega_{i})D[\tilde{a}^{\dagger}\tilde{b}_{i}^{n}]
	+  f_{q}(\omega_{a}+n\omega_{i})D[\tilde{a}\tilde{b}_{i}^{n}]
	+  f_{q}(-\omega_{a}-n\omega_{i})D[\tilde{a}^{\dagger}\tilde{b}_{i}^{\dagger n}]\big\},
	+\alpha^{3}\bigg\{f_{q}(\omega_{a}-\sum_{i}\omega_{i})D[\tilde{a}\tilde{b}_{1}^{\dagger}\tilde{b}^{\dagger}_{2}]
\nonumber \\ &&+  f_{q}(-\omega_{a}+\sum_{i}\omega_{i})D[\tilde{a}^{\dagger}\tilde{b}_{1}\tilde{b}_{2}] + f_{q}(\omega_{a}+\sum_{i}\omega_{i})D[\tilde{a}\tilde{b}_{1}\tilde{b}_{2}]+f_{q}(-\omega_{a}-\sum_{i}\omega_{i})D[\tilde{a}^{\dagger}\tilde{b}_{1}^{\dagger}\tilde{b}^{\dagger}_{2}] + f_{q}(\omega_{a}-\omega_{1}+\omega_{2})D[\tilde{a}\tilde{b}_{1}^{\dagger}\tilde{b}_{2}] 
\nonumber\\&&+f_{q}(-\omega_{a}+\omega_{1}-\omega_{2})D[\tilde{a}^{\dagger}\tilde{b}_{1}\tilde{b}^{\dagger}_{2}] + f_{q}(\omega_{a}+\omega_{1}-\omega_{2})D[\tilde{a}\tilde{b}_{1}\tilde{b}^{\dagger}_{2}]+f_{q}(-\omega_{a}-\omega_{1}+\omega_{2})D[\tilde{a}^{\dagger}\tilde{b}^{\dagger}_{1}\tilde{b}_{2}]\bigg\}, \nonumber\\
	\tilde{\mathcal{L}}_{i}\tilde{\rho} &=& f_{i}(\omega_{i})D[\tilde{b}_{i}]
	+ f_{i}(-\omega_{i})D[\tilde{b}_{i}^{\dagger}]. \label{eq:L_M}
\end{eqnarray}
\end{widetext}
We note that in the derivation of the master equation, we ignored all higher-order terms $\mathcal{O}(\alpha^4)$ and assumed $\alpha = \alpha_{i}$ for simplicity. The Lindblad dissipator $D[\hat{o}]$ is defined by
\begin{equation}
D[\tilde{o}] = \frac{1}{2}(2\tilde{o}\tilde{\rho}\tilde{o}^{\dagger}-\tilde{\rho}\tilde{o}^{\dagger}\tilde{o}-\tilde{o}^{\dagger}\tilde{o}\tilde{\rho}).
\end{equation}
The spectral response function of the baths is denoted by $\tilde{f}(\omega)$. In the rest of the paper,  unless otherwise mentioned, we consider one-dimensional Ohmic spectral densities for all the thermal baths, given by
\begin{eqnarray}\label{eq:SRF}
f_{E}(\omega)=
\begin{cases}
\omega\gamma_{E}[1 + \bar{n}_{E}(\omega)], &\omega> 0, \\
\abs{\omega}\gamma_{E}\bar{n}_{E}(\abs{\omega}), &\omega< 0, 
\end{cases}
\end{eqnarray}
and $f_{E}(0)=0$ for Ohmic spectral densities of the baths.
Here, the system-bath coupling strength is denoted by $\gamma_{E}$, and the mean number of quanta in each thermal bath is given by (we take the Boltzmann constant $k_{B}=1$)
\begin{equation}
\bar{n}_{E}(\omega)= \frac{1}{e^{\omega/T_{E}}-1}.
\end{equation}

\section{Results and discussion}\label{sec:Results}
In this section, we shall investigate the entanglement between the resonators in the model presented in Fig.~\ref{fig:fig1}. This scheme employs bath spectrum filtering, an instance of reservoir engineering, to generate entanglement between the resonators. For the numerical analysis, we consider the parameters of an electro-mechanical system~\cite{LaHaye2009, Massel2012}: $\omega_{a}=2\pi\times 10$ GHz, $\omega_{i}=2\pi\times 10$ MHz, $\gamma_{h}=\gamma_{c}=2\pi\times 5$ MHz, $\gamma_{i}=2\pi\times 100$ Hz, and $g_{i}=2\pi\times 500$ KHz, unless otherwise stated.

In the rest of the work, we consider TLS ancilla to illustrate our proposed scheme. If a TLS or a harmonic oscillator interacts with the resonators via {\it{energy-field}} coupling [Eq.~(\ref{eq:sysHam})], an external coherent control drive of frequency $\omega_{c}=\abs{\omega_{a}\pm n\omega_{i}}$ can induce sideband transitions of the order $n$. For instance, a control drive resonant with the lower first sideband can induce sideband cooling of the resonator~\cite{Teufel2011}. If the drive is resonant with the lower second sideband, it may produce antibunching of the resonator field~\cite{PhysRevA.85.051803}.
 In a complementary approach, if an incoherent suitably engineered thermal bath, drives the subsystem $A$ at the lower first sideband, it leads to cooling of resonators, colloquially referred to as ``cooling by heating". In this case, $D[\tilde{a}^{\dagger}\tilde{b}_{i}]$ dominates the dissipative dynamics of the system~\cite{Naseem2021}.
Similarly, an incoherent drive at the lower second sideband results in the antibunching of the phonon field, in which $D[\tilde{a}^{\dagger}\tilde{b}^2_{i}]$ is the dominant source of dissipation~\cite{PhysRevA.105.012201}.
Remarkably, the rate equations derived in both coherent and incoherently driven $A-R_{i}$ system shares a similar mathematical structure~\cite{Naseem2021, PhysRevA.105.012201}.

By noting this analogy, one may consider the incoherent coupling counterpart of the two-mode squeezing interaction, i.e., $D[\tilde{a}^{\dagger}\tilde{b}_{1}\tilde{b}_{2}]$ ($D[\tilde{a}\tilde{b}^{\dagger}_{1}\tilde{b}^{\dagger}_{2}]$), may lead to entanglement between the  resonators $R_{i}$. However, we aim to entangle modes $\hat{b}_{i}$ and not necessarily the transformed modes $\tilde{b}_{i}$. Hence, $D[\tilde{a}^{\dagger}\tilde{b}_{1}\tilde{b}_{2}]$ in the {\it{local}} basis of the resonators, after taking a trace over subsystem $A$, takes the form
\begin{equation}\label{eq:EntDiss}
D_\text{ent} = \text{Tr}_{A}[\hat{u}^{\dagger}D[\tilde{a}^{\dagger}\tilde{b}_{1}\tilde{b}_{2}]\hat{u}]= \langle\hat{n}_{a}+1\rangle D[e^{\hat{c}-\hat{c}^{\dagger}}(\hat{d}-\hat{c})],
\end{equation}
with 
\begin{eqnarray}
\hat{c} &:=& \alpha(\hat{b}_{1}+\hat{b}_{2}), \nonumber \\
\hat{d} &:=& \hat{b}_{1}\hat{b}_{2} + \alpha^{2}.
\end{eqnarray}
In a previous study~\cite{Arenz_2013}, it was shown that a nonclassical entangled state of two microwave cavity fields could be engineered via a collision model. In this scheme, carefully prepared three-level atoms with coherent external control drives, randomly
interact with the cavity fields. This gives an effective dissipative dynamics of the cavity fields, governed by the Liouville super-operator~\cite{Arenz_2013} 
\begin{equation}\label{eq:VitaliDiss}
\mathcal{L}_\text{eff}\hat{\rho} = \kappa_{c}D[\hat{c}_{-}]+\kappa_{d}D[\hat{d}_{-}].
\end{equation}
Here, $\kappa_{c, d}$ are the dissipation rates and these depend on the specific system parameters.
$\hat{c}_{-}$ and $\hat{d}_{-}$ are the joint jump operator for the cavity field and given by~\cite{Arenz_2013} 
\begin{eqnarray}\label{eq:vitoperators}
\hat{c}_{-} &=& \frac{1}{\sqrt{2}}(\hat{b}_{1}-\hat{b}_{2}), \\ \nonumber
\hat{d}_{-} &=& \frac{1}{2}\hat{b}_{1}\hat{b}_{2} - \beta^2,
\end{eqnarray}
here $\beta$ represents the complex amplitude of the coherent states of both cavities.
Eq.~(\ref{eq:VitaliDiss}) shows that $D_\text{ent}$ may lead to an entangled state of the uncoupled resonators $R_{i}$. However, there are additional terms in $D_\text{ent}$ compared with Eq.~(\ref{eq:VitaliDiss}). To see the effect of these terms on entanglement generation, we consider a hypothetical setting where two uncoupled resonators evolve under the dissipative dynamics governed by $D_\text{ent}$. Per se $D_\text{ent}$ acts alone, we compare the entanglement results of this setting with that of Eq.~(\ref{eq:VitaliDiss}).

\begin{figure}[t!]
  \centering
  \includegraphics[scale=0.52]{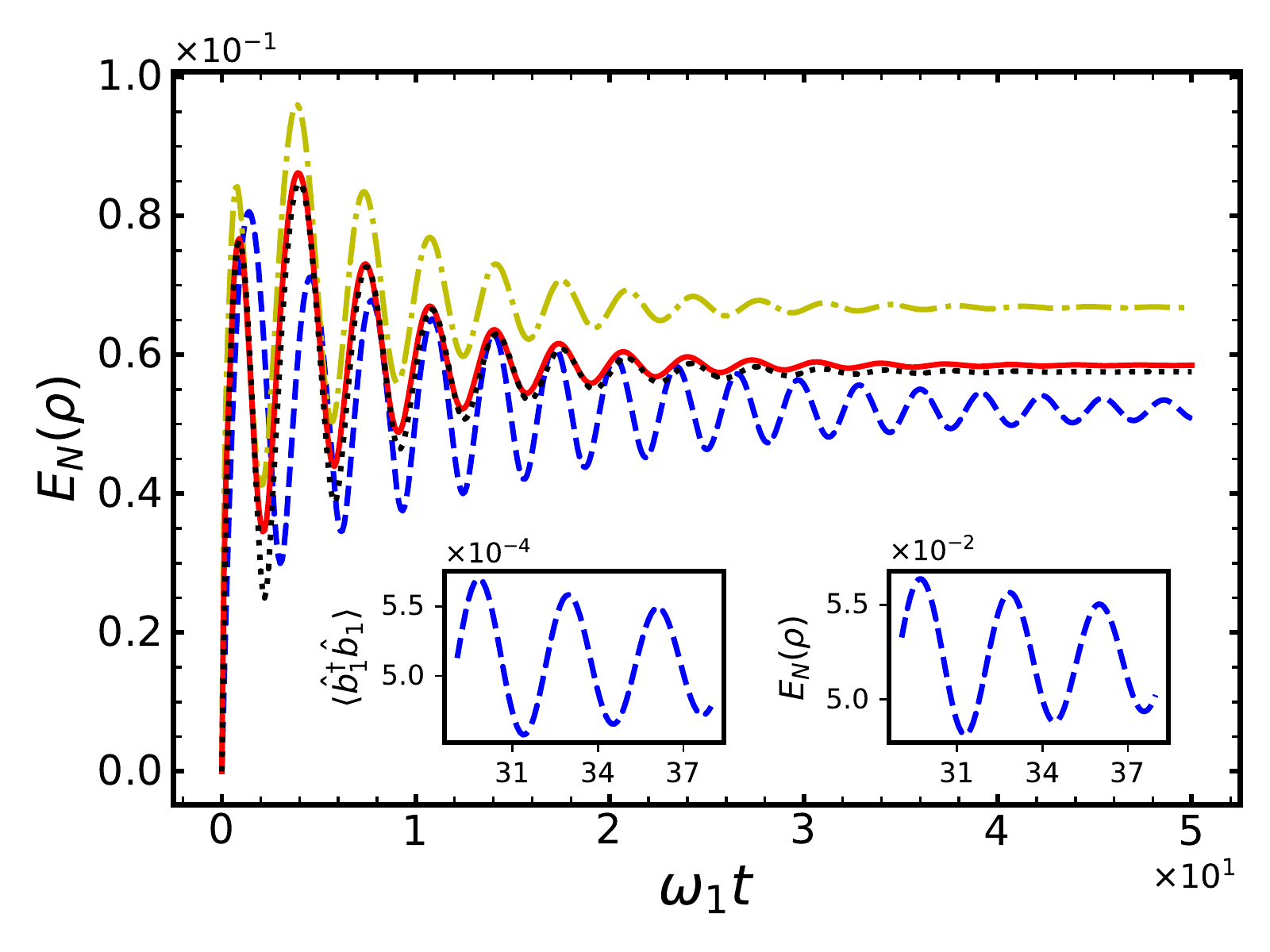}
  \caption{Physical mechanism of entanglement generation. The logarithmic negativity $E_{N}(\rho)$ as a function of scaled time $\omega_{1}t$. The blue dashed curve is obtained by the numerical integration of Eq.~(\ref{eq:VitaliDiss}), proposed in~\cite{Arenz_2013} for generating a two-mode entangled state of optical resonators. The black dotted curve denotes $E_{N}(\rho)$ for a hypothetical case in which two uncoupled resonators evolve under dissipative dynamics governed by Eq.~(\ref{eq:EntDiss}). The yellow dot-dashed and the red solid lines show $E_{N}(\rho)$ by keeping the first order and second-order terms in $\alpha$, respectively, and by ignoring all higher-order terms in Eq.~(\ref{eq:EntDiss}). The right inset shows a part of the blue dashed curve for $E_{N}(\rho)$ . Finally, in the left inset, the mean number of excitation $\langle \hat{b}^{\dagger}_{1}\hat{b}_{1}\rangle$ in the resonator $R_{1}$ is plotted as a function of scaled time $\omega_{1} t$. All these curves show an almost periodic behavior with a period $T\approx {2\pi}/{\sum_{i}\omega_{i}}$. Initially, the resonators are prepared in a product of ground-states, $\hat{\rho}(0)=\hat{\rho}_{1}\otimes\hat{\rho}_{2}=\ket{0_{1}}\bra{0_{1}}\otimes\ket{0_{2}}\bra{0_{2}}$. In the numerical simulations, we use the Hilbert space with ten photon states for each resonator, and it is verified that the increase in the Hilbert space dimensions does not affect the results.
   Parameters: $\omega_{i}=1$, $\alpha_{i}=0.2$, and $\kappa_{c}=\kappa_{d}=0.1$.}
  \label{fig:fig2}
\end{figure}

The initial state of the resonators is considered to be a separable ground-state, $\hat{\rho}(0)=\hat{\rho}_{1}\otimes\hat{\rho}_{2}=\ket{0_{1}}\bra{0_{1}}\otimes\ket{0_{2}}\bra{0_{2}}$. Other initial states such as thermal state can also be considered; however, entanglement remains small or insignificant for states other than the ground-states of the resonators~\cite{Arenz_2013}. To quantify the entanglement between resonators; we use logarithmic negativity~\cite{PhysRevLett.95.090503}, which is given by
\begin{equation}\label{eq:negativity}
E_{N}(\hat{\rho}) = \text{log}_{2}\|\hat{\rho}^{\Gamma_{2}}\|.
\end{equation}
Here, $\hat{\rho}^{\Gamma_{2}}$ is the partial transpose of the joint density matrix of resonators with respect to $R_{2}$, and $\|.\|$ denotes the trace norm of $\hat{\rho}^{\Gamma_{2}}$. “In contrast to systems in pure states, bipartite entanglement cannot be uniquely and easily defined in terms of the entropy for systems in mixed states. Among different witnesses and bounds used to examine quantum entanglement in mixed states~\cite{DBLP, RevModPhys.81.865}, (logarithmic) negativity is commonly used for characterization of mixed state entanglement in quantum technology and condensed matter applications~\cite{RevModPhys.81.865, PhysRevA.65.032314, Lanyon2017}. The main motivations behind its choice are that it is computationally tractable, it carries a clear operational meaning being an upper bound for distillable entanglement, and it has direct relevance to experimental measurements of mixed state quantum entanglement~\cite{PhysRevLett.125.200501}. Our approach relies on bath induced quantum correlations between subsystems, and hence it is an open quantum system in mixed states, and motivated by these reasons we also choose negativity to quantify quantum entanglement in our case.

The comparison of $E_{N}(\hat{\rho})$ for two uncoupled oscillators undergoing dissipative dynamics governed by Eq.~(\ref{eq:EntDiss}) and Eq.~(\ref{eq:VitaliDiss}) is shown in Fig.~\ref{fig:fig2}. All the parameters are considered identical in both cases, including dissipation rates. In the numerical results, we use the Hilbert space size with ten photon states for each resonator and verified that the results do not change by the increase in the Hilbert space size. The result shows that for the considered parameters, entanglement becomes more pronounced in the case of $D_\text{ent}$. This result indicates that additional terms in $D_\text{ent}$ do not destroy the entanglement between the resonators. In the transient regime, both entanglement $E_{N}(\hat{\rho})$ and populations $\langle \hat{b}^{\dagger}_{1}\hat{b}_{1}\rangle$ show similar oscillatory behavior with a period $T=2\pi/\sum_{i}\omega_{i}$, and these oscillations vanish in the long time limit. 
The numerical results in Fig.~\ref{fig:fig2} show that the populations govern the dynamics of the Logarithmic negativity. As $E_{N}(\hat{\rho})$ is found to be maximum for the maximum values of populations and vice versa. 
Similar behavior has previously been reported for linearly coupled oscillators evolving under nonlinear dissipation~\cite{PhysRevA.88.022309}.  

It is worth mentioning that we do not assume two-mode squeezing-like interaction between the resonators. In our model, the resonators are uncoupled. The microscopic derivation of the master equation (Eq.~(9)) results in the bath-mediated incoherent interactions between the resonators, among which two-mode squeezing-like interactions are also present. These incoherent interactions are induced by the thermal environment and should not be mixed with the two-mode squeezing interactions that appear in the system's Hamiltonian. We like to emphasize that it is not the two-mode squeezing-like interactions per se that lead to entanglement between the resonators. The transformation of these dissipators in the ``local" basis of the resonators yields required dissipation (Eq.~(13)). Hence, only two-mode squeezing-like Lindblad dissipators are not sufficient to create entanglement between the resonators.

\begin{figure}[t!]
  \centering
  \includegraphics[scale=0.52]{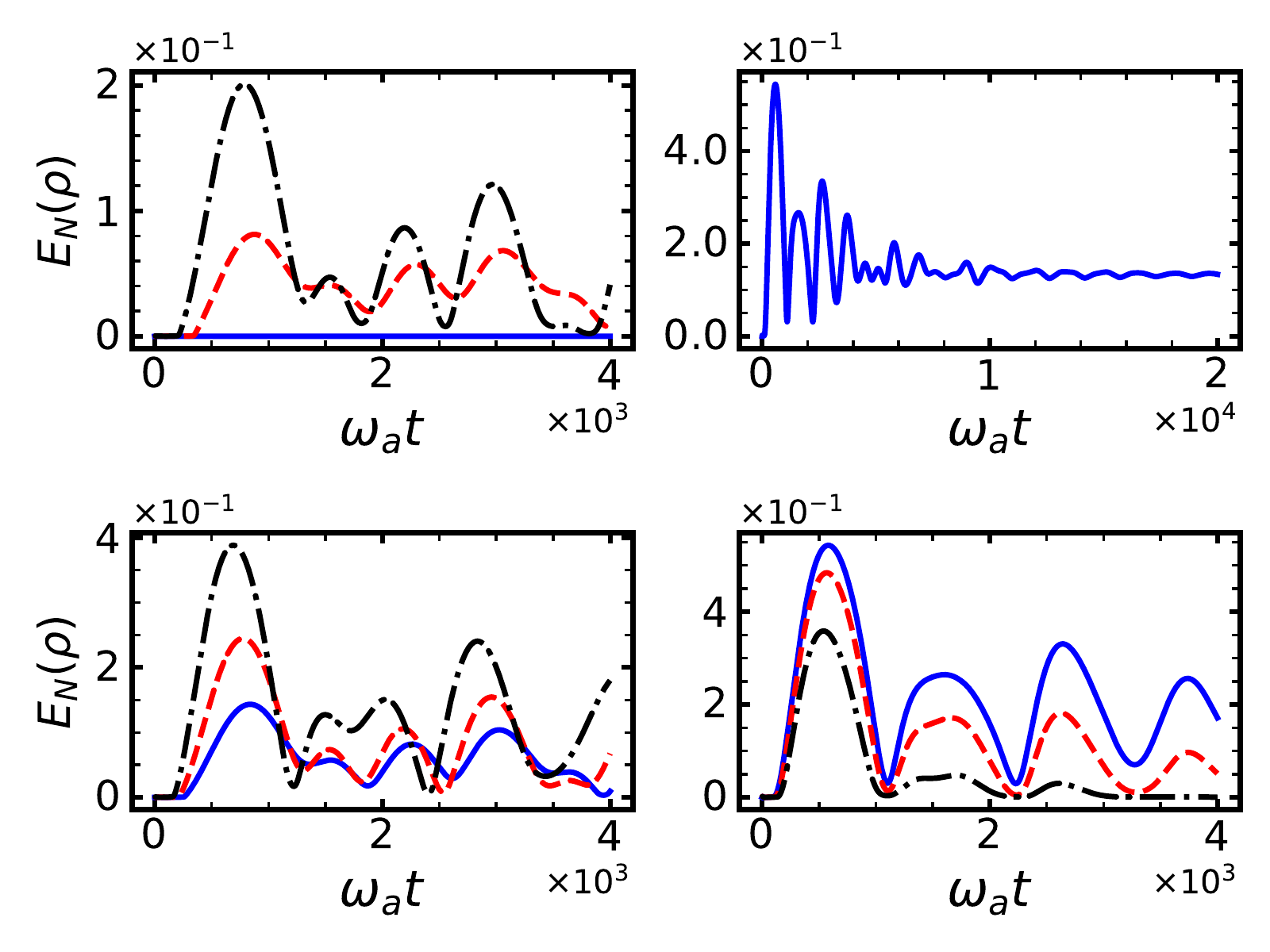}
  \caption{Entanglement quantification of resonators with different frequencies. Logarithmic negativity $E_{N}(\rho)$ as a function of scaled time $\omega_{a}t$. The results are obtained by the numerical integration of Eq.~(\ref{eq:filtMEM}) and then using the inverse transformation [Eq.~(\ref{eq:transform})] to get $\hat{\rho}$ in the {\it{local}} basis. Finally, $E_{N}(\rho)$ is computed numerically using Eq.~(\ref{eq:negativity}). Initial state of the system is considered a pure separable state of the subsystems, $\hat{\rho}(o)=\hat{\rho}_{a}\otimes\hat{\rho}_{1}\otimes\hat{\rho}_{2}=
  \ket{g_{a}}\bra{g_{a}}\otimes\ket{0_{1}}\bra{0_{1}}\otimes\ket{0_{2}}\bra{0_{2}}$, here $\ket{g_{a}}$ represents the ground state of the TLS. Parameters: $\omega_{a}=2\pi\times 10$ GHz, $\omega_{1}=2\omega_{2}= 2\pi\times 5$ MHz, $\gamma_{h}=\gamma_{c}=2\pi\times 500$ KHz, $\gamma_{i}=2\pi\times 100$ Hz, and $T_{c}=65$ mK. All the parameters are scaled with $\omega_{a}$ in the numerical results. {\bf(a)} Blue solid, red dashed, and black dot-dashed curves are for $T_{h}=1, 70, 300$ K, respectively. Here, $T_{i}=0.1$ K and  $\alpha_{i}=0.1$. {\bf(b)} Long time entanglement $E_{N}(\rho)$ for $\alpha_{i}=0.2$, $T_{h}=300$ K and $T_{i}=0.05$ K. {\bf(c)} Blue solid, red dashed, and black dot-dashed curves are for $\alpha_{i}=0.05, 0.1, 0.2$, respectively. $T_{h}=300$ K and $T_{i}=0.1$ K. {\bf(d)} Blue solid, red dashed, and black dot-dashed curves are for $T_{i}=0.1, 0.15, 0.2$ K, respectively. $T_{h}=300$ K and $\alpha_{i}=0.2$.}
  \label{fig:fig3}
\end{figure}

It remains to show that, in our scheme, $D_\text{ent}$ is the dominant source of dissipation in the
the dynamics of the system $\hat{H}_{S}$ given in Eq.~(\ref{eq:sysHam}). To this end, we employ bath spectrum filtering to remove the unwanted dissipative channels. We consider the filtered bath spectra of the hot and cold baths shown in Fig.~\ref{fig:fig1}(b) and given by~\cite{doi:10.1080/09500349414550381, Ghosh12156, GELBWASERKLIMOVSKY2015329, Naseem_2020, PhysRevResearch.2.033285, Naseem2021}
\begin{equation}\label{eq:filter}
\tilde{f}_{q}(\omega) = \frac{\kappa_{q}}{\pi} \frac{(\pi f_{q}(\omega))^2}{(\omega -(\omega^{r}_{q}+\Delta_{q}(\omega))^2+(\pi f_{q}(\omega))^2}.
\end{equation} 
Here $\kappa_{q}$ is the coupling strength between subsystem $A$ and the bath filter, and $\omega^{r}_{q}$ is the resonance frequency of the bath spectrum. The bath induced Lamb shift is given by
\begin{equation}
\Delta_{q}(\omega) = P \int^{\infty}_{o} d\omega^{'} \frac{f_{q}(\omega^{'})}{\omega-\omega^{'}},
\end{equation}
here $P$ denotes the principal value. The bath modes closer to the resonance frequency $\omega^{r}_{q}$ are more strongly coupled to the system.

We shall discuss the entanglement generation in two different situations, depending upon the frequencies of the resonators.
\begin{figure}[t!]
  \centering
  \includegraphics[scale=0.52]{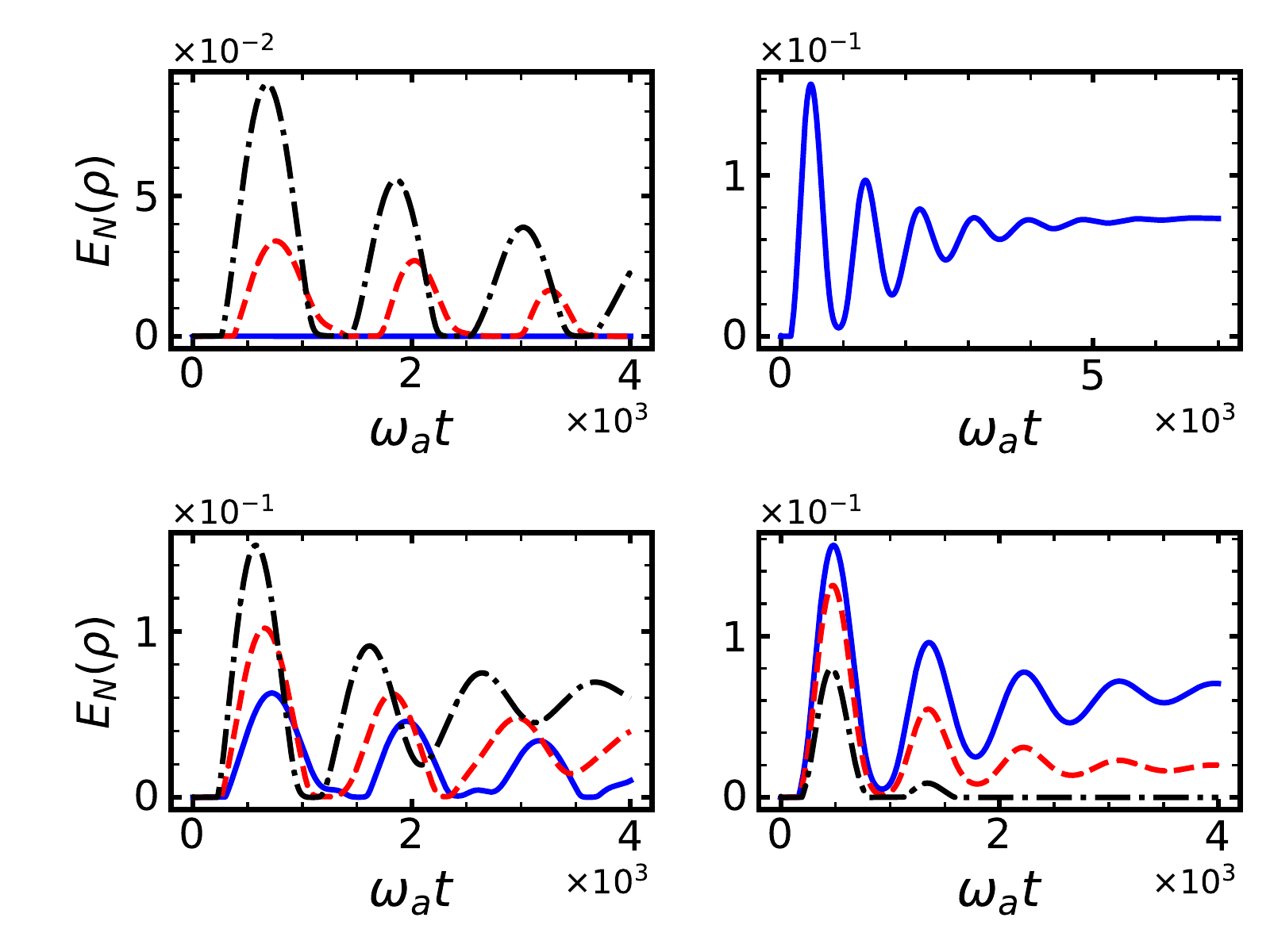}
  \caption{Entanglement quantification of resonators with same frequencies. Logarithmic negativity $E_{N}(\rho)$ as a function of scaled time $\omega_{a}t$. The results are obtained by the numerical integration of Eq.~(\ref{eq:filtMEM2}) and then using the inverse transformation [Eq.~(\ref{eq:transform})] to get $\hat{\rho}$ in the {\it{local}} basis. Finally, $E_{N}(\rho)$ is computed numerically using Eq.~(\ref{eq:negativity}). Initial state of the system is considered a pure separable state of the subsystems, $\hat{\rho}(o)=\hat{\rho}_{a}\otimes\hat{\rho}_{1}\otimes\hat{\rho}_{2}=
  \ket{g_{a}}\bra{g_{a}}\otimes\ket{0_{1}}\bra{0_{1}}\otimes\ket{0_{2}}\bra{0_{2}}$, here $\ket{g_{a}}$ represents the ground state of the TLS.
  Parameters: $\omega_{a}=2\pi\times 10$ GHz, $\omega_{1}=\omega_{2}= 2\pi\times 5$ MHz, $\gamma_{h}=\gamma_{c}=2\pi\times 500$ KHz, $\gamma_{i}=2\pi\times 100$ Hz, and $T_{c}=65$ mK. All the parameters are scaled with $\omega_{a}$ in the numerical results. {\bf(a)} Blue solid, red dashed, and black dot-dashed curves are for $T_{h}=1, 70, 300$ K, respectively. Here, $T_{i}=0.1$ K and  $\alpha_{i}=0.1$. {\bf(b)} Long time entanglement $E_{N}(\rho)$ for $\alpha_{i}=0.2$, $T_{h}=300$ K and $T_{i}=0.05$ K. {\bf(c)} Blue solid, red dashed, and black dot-dashed curves are for $\alpha_{i}=0.05, 0.1, 0.2$, respectively. $T_{h}=300$ K and $T_{i}=0.1$. {\bf(d)} Blue solid, red dashed, and black dot-dashed curves are for $T_{i}=0.1, 0.15, 0.2$ K, respectively. $T_{h}=300$ K and $\alpha_{i}=0.2$. All the curves show an almost periodic behavior of $E_{N}(\rho)$ with a period of $T\approx {4\pi}/{\sum_{i}\omega_{i}}$.}
  \label{fig:fig4}
\end{figure}

\subsection{Non-degenerate resonators}\label{subsec:NDR}
First we discuss the case in which difference between the frequencies of resonators is large. If we consider the resonance frequency of the hot bath filter $\omega^{r}_{h}=\omega_{a}-\sum_{i}\omega_{i}$ and of the cold bath filter $\omega^{r}_{c}=\omega_{a}$ in Eq.~(\ref{eq:filter}) (shown in Fig.~\ref{fig:fig1}(b)), this choice leads to Liouville super-operators of the hot and cold bath of the form
\begin{eqnarray}
\tilde{ \mathcal{L}}^{d}_{c}\tilde{\rho} &=&  \tilde{f}_{c}(\omega_{a})\{D[\tilde{a}]+\alpha^{3}\sum_{i}D[\tilde{a}\tilde{b}_{i}^{\dagger}\tilde{b}_{i}]\}
\nonumber \\
&&+ \tilde{f}_{c}(-\omega_{a})\{D[\tilde{a}^{\dagger}]  + \sum_{i}\alpha^{3}D[\tilde{a}^{\dagger}\tilde{b}_{i}^{\dagger}\tilde{b}_{i}]\},\nonumber \\ 	
\tilde{\mathcal{L}}^{d}_{h}\tilde{\rho}	&=& \alpha^3\big\{\tilde{f}_{h}(\omega_{-})D[\tilde{a}\tilde{b}_{1}^{\dagger}\tilde{b}_{2}^{\dagger}]
	+ \tilde{f}_{h}(-\omega_{-})D[\tilde{a}^{\dagger}\tilde{b}_{1}\tilde{b}_{2}]\big\},\nonumber \\ 
	\tilde{\mathcal{L}}_{i}\tilde{\rho} &=& f_{i}(\omega_{i})D[\tilde{b}_{i}]
	+ f_{i}(-\omega_{i})D[\tilde{b}_{i}^{\dagger}],\label{eq:filtMEM}
\end{eqnarray}
with $\omega_{-}=\omega_{a}-\sum_{i}\omega_{i}$ and $\tilde{\mathcal{L}}_{i}$ remains unchanged. For $T_{h}\gg T_{i}>T_{c}$ and $\gamma_{i}\bar{n}_{i}\ll \alpha^{3}\tilde{f}_{h}(\omega_{-})$, $D_\text{ent}$ is the dominant source of dissipation in the system. This leads to considerable degree of entanglement between the resonators $R_{i}$. In rest of the work, resonators $R_{i}$ are considered in an initial separable state of their ground states. This initial state could be achieved without the need for external control drive in our scheme. Simultaneous ground state cooling of the resonators can be achieved in the limit $T_{h}\gg T_{i}>T_{c}$ by tuning the resonance frequency of the hot bath filter at the first lower sideband [see Eq.~(\ref{eq:filter})]~\cite{Naseem2021}. Then one switch the resonance frequency to lower second order sideband, soon after resonators cool to their ground states.

Time dependence of logarithmic negativity $E_{N}(\hat{\rho})$, in case of TLS$-R_{i}$ system is shown in Fig.~\ref{fig:fig3}.
 For  $T_{h}\sim T_{i}>T_{c}$, there is no entanglement between resonators $R_{i}$ because $\tilde{\mathcal{L}}_{i}$ dominates the dissipation dynamics in this limit. In case of $T_{h}\gg T_{i}>T_{c}$, entanglement between the resonators emerges and it increases with the higher values of $T_{h}$, shown in Fig.~\ref{fig:fig3}(a). Entanglement $E_{N}(\hat{\rho})$ between the resonators in the long-time limit is shown in Fig.~\ref{fig:fig3}(b). 
  Compared with Fig.~\ref{fig:fig2} results, there is no periodic behavior of entanglement in Fig.~\ref{fig:fig3}. This is because in  Fig.~\ref{fig:fig3}, we consider non-degenerate resonators $R_{i}$, which gives non-synchronized dynamics of the mean number of excitations $\hat{n}_{i}=\langle \hat{b}^{\dagger}_{i}\hat{b}_{i}\rangle$. As shown in Fig.~\ref{fig:fig2}, $\hat{n}_{i}$ governs the dynamics of $E_{N}(\hat{\rho})$, hence, non-synchronized dynamics dynamics of $\hat{n}_{i}$ suggests the absence of periodicity in $E_{N}(\hat{\rho})$. The increase in entanglement with the increase in coupling strength $g_{i}$ between TLS and resonators is presented in Fig.~\ref{fig:fig3}(c). We check the robustness of entanglement between the resonators against thermal vibrations in Fig.~\ref{fig:fig3}(d).
 
  These results show that the early time entanglement is reasonably robust against resonators environment temperature provided sufficiently large $\alpha$ and $T_{h}\gg T_{i}>T_{c}$. However, for large bath temperatures $T_{i}\gg\omega_{a}$, entanglement shows damped oscillations and vanishes in the long-time limit. We note that for higher bath temperatures $T_{i}$, larger values of $\alpha$ and very high hot bath temperature $T_{h}$ are required to generate long-time entanglement between the resonators. This is because $D[\tilde{a}^{\dagger}\tilde{b}_{1}\tilde{b}_{2}]$ is responsible for both  cooling and creating entanglement. For $T_{i}\gg\omega_{a}$, stronger coupling strength $g_{i}$ and very high-temperature $T_{h}$ are required to cool the resonators. The use of a third common bath to enhance the entanglement by cooling the system is consistent with previously reported results~\cite{Man_2019}. In this scheme, entanglement enhancement is shown by cooling a pair of qubits via a shared bath. Alternatively, the hot bath can be replaced by a squeezed thermal bath or reservoir with a negative temperature to enhance the cooling and entanglement between resonators~\cite{PhysRevLett.120.063604}.  

\subsection{Degenerate resonators}\label{subsec:DR}
For the case of degenerate resonators $\omega_{1}=\omega_{2}=\Omega$, if we select the resonance frequency of the hot bath $\omega^{r}_{h}=\omega_{a}-2\Omega$, and of the cold bath $\omega^{r}_{c}=\omega_{a}$ in Eq.~(\ref{eq:filter}), the master equation (\ref{eq:L_L}) reduces to
\begin{eqnarray}
\tilde{ \mathcal{L}}^{s}_{c}\tilde{\rho} &=&  \tilde{\mathcal{L}}^{d}_{c}\tilde{\rho} + \alpha^{3}\tilde{f}_{c}(\omega_{a})\{D[\tilde{a}\tilde{b}^{\dagger}_{1}\tilde{b}_{2}] + D[\tilde{a}\tilde{b}_{1}\tilde{b}^{\dagger}_{2}]\}\nonumber \\  &&\qquad + \alpha^{3}\tilde{f}_{c}(-\omega_{a})\{D[\tilde{a}^{\dagger}\tilde{b}_{1}\tilde{b}^{\dagger}_{2}] + D[\tilde{a}^{\dagger}\tilde{b}^{\dagger}_{1}\tilde{b}_{2}]\}
,\nonumber \\ 	
\tilde{\mathcal{L}}^{s}_{h}\tilde{\rho}	&=& \tilde{\mathcal{L}}^{d}_{h}\tilde{\rho}+\alpha^{3}\{\tilde{f}_{h}(\omega_{a}-2\Omega)\sum_{i}D[\tilde{a}\tilde{b}_{i}^{\dagger 2}] \nonumber \\ 
&&\qquad\qquad +\tilde{f}_{h}(-\omega_{a}+2\Omega)\sum_{i}D[\tilde{a}^{\dagger}\tilde{b}^{2}_{i}]\},\nonumber \\ 
	\tilde{\mathcal{L}}_{i}\tilde{\rho} &=& f_{i}(\Omega)D[\tilde{b}_{i}]
	+ f_{i}(-\Omega)D[\tilde{b}_{i}^{\dagger}].\label{eq:filtMEM2}
\end{eqnarray}
There are additional terms in Eq.~(\ref{eq:filtMEM2}) compared with Eq.~(\ref{eq:filtMEM}). We consider 
$T_{c}\ll 1$ K in our results, hence, the impact of additional terms of $\tilde{ \mathcal{L}}^{s}_{c}\tilde{\rho}$ on entanglement generation is not significant. Contrary, the second and third terms of $\tilde{ \mathcal{L}}^{s}_{c}\tilde{\rho}$ dissipation rates depends on $T_{h}$, consequently, have
non-negligible negative effect on the entanglement. The results in Fig.~\ref{fig:fig4} confirms the decrease in the entanglement for the degenerate resonators due to the presence of additional terms in Eq.~(\ref{eq:filtMEM2}). In Fig.~\ref{fig:fig4}, boAccordingly, the cooling of the mechanical resonators is accompanied by entanglement generation (). This analysis is confirmed by the numerical results presented in Fig. 5, which show that the cooling and entanglement co-occur in our scheme. For the entanglement generation, the ground state of the resonators is an optimal choice. Fig. 5 shows that the thermal state with sufficiently low excitations ni can also be considered. th resonators are identical, which leads to synchronized dynamics of the mean excitations $\langle\tilde{b}^{\dagger}_{i}\tilde{b}_{i}\rangle$. In addition, both populations and entanglement show periodic oscillations with a period of $T=2\pi/\Omega$. Note that the early time entanglement behavior is monotonic as a function of the hot bath temperature, as shown in Fig.~\ref{fig:fig4}(a). This is in contrast to previous works~\cite{PhysRevLett.88.197901, Bohr_Brask_2015}, where entanglement increases to a maximum value for the optimal value of driving heat bath temperature, and then decreases to zero for higher temperatures.

Our proposal can be experimentally realized using an electro-mechanical system~\cite{Massel2012} or circuit QED setup~\cite{PhysRevLett.120.227702}. The system consists of a superconducting microwave resonator (SR), or a qubit of frequency $\omega_{a}$, which is simultaneously coupled to two micromechanical resonators~\cite{Massel2012}, or microwave resonators~\cite{PhysRevLett.115.203601, PhysRevLett.120.227702} of frequency $\omega_{i}$. Spectrally filtered hot and cold baths at temperatures $T_{h}$ and $T_{c}$ drive the SR, respectively. In addition, two microwave resonators designed at frequencies $\omega_{-}=\omega_{a}-2\sum_{i}\omega_{i}$, and $\omega_{a}$ couple to the SR for the realization of the hot and cold baths spectrum filtering, respectively. For the implementation of the thermal baths, each of these microwave resonators is coupled to a copper thin-film resistor~\cite{Senior2020}. The spectral density of the bath provided by these resistors can be modeled by Ohmic spectral density~\cite{Zueco_2008}. In the bath spectrum filtering, it is possible to map the microwave resonators (coupled to the SR or qubit) and the baths to a structured environment of non-interacting harmonic oscillators with an effective spectral density given by Lorentzian function~\cite{PhysRevB.101.245415}. The judicious choice of the microwave resonator frequencies completely suppresses or weakens the undesired bath-induced transitions.

\section{Conclusions}\label{sec:conc}
We proposed and analyzed an autonomous quantum thermal machine to generate transient and steady-state entanglement of two uncoupled macroscopic mechanical resonators. Unlike previous proposals on reservoir engineering, our quantum entanglement engine does not rely on external control or work input to generate entanglement between uncoupled macroscopic mechanical resonators. Entanglement between two uncoupled resonators is created by coupling them to a common incoherently driven two-level system or a harmonic oscillator. We proposed bath spectrum filtering to engineer desirable incoherent interaction with the thermal baths to create entanglement, which is a critical step in our scheme. Our numerical results showed a significantly large amount of entanglement established between the resonators in the absence of thermal equilibrium and even in the presence of the local dissipation of the resonators. 
In addition, the amount of early
time entanglement showed a monotonic increase with the increase in a temperature gradient. To the best of our knowledge, there are no reported results of steady-state entanglement generation between uncoupled mechanical resonators using temperature gradient. If we consider circuit QED implementation of our scheme, then entanglement generation may be compared with Ref.~\cite{PhysRevLett.88.197901} in which entanglement between two optical cavities is investigated in a similar setting. In our system, entanglement is almost two orders of magnitude larger than reported in Ref.~\cite{PhysRevLett.88.197901}.

 The periodic oscillations in the entanglement are reported for the resonators of the same frequency. For the non-degenerate case, the periodicity breaks down due to the non-synchronized dynamics of the resonators.
For the experimental realizations, several physical systems are conceivable such as optomechanical system~\cite{RevModPhys.86.1391}, circuit QED~\cite{PhysRevLett.120.227702, Bothner2021}, and electro-mechanical systems~\cite{LaHaye2009}.
\begin{figure}[t!]
  \centering
  \includegraphics[scale=0.52]{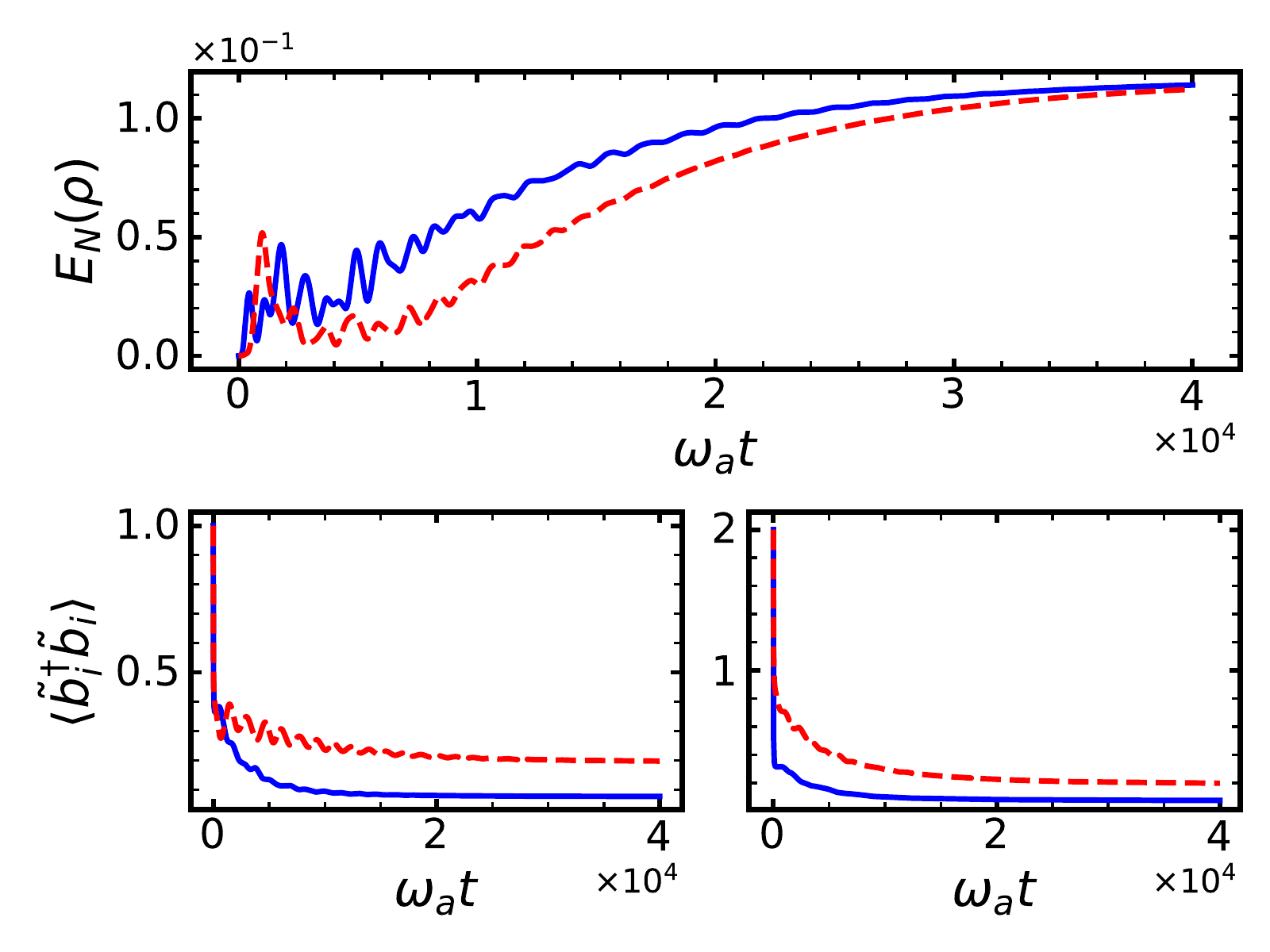}
  \caption{{\bf(a)} Entanglement quantification $E_{N}(\rho)$ of the resonators with different initial thermal states as a function of scaled time $\omega_{a}t$. The solid and dashed lines represent the initial thermal state of the resonators with the mean excitation $\bar{n}_{i}=1, 2$, respectively. The results are obtained by the numerical integration of Eq.~(\ref{eq:filtMEM}) and then using the inverse transformation [Eq.~(\ref{eq:transform})] to get $\hat{\rho}$ in the {\it{local}} basis. Finally, $E_{N}(\rho)$ is computed numerically using Eq.~(\ref{eq:negativity}). 
 {\bf(b)}, and {\bf(c)} show the mean phonon number $\langle \tilde{b}^{\dagger}_{i}\tilde{b}_{i}\rangle$ for initial mean excitation $\bar{n}_{i}=1, 2$, as a function of scaled time $\omega_{a}t$, respectively. The solid and dashed lines are for the resonaotrs $R_{1}$ and $R_{2}$, respectively.
 Parameters: $\omega_{a}=2\pi\times 10$ GHz, $\omega_{1}=2\omega_{2}= 2\pi\times 5$ MHz, $\gamma_{h}=\gamma_{c}=2\pi\times 500$ KHz, $\gamma_{i}=2\pi\times 100$ Hz,  $\alpha_{i}=0.2$, $T_{h}=300$ K, $T_{c}=65$ mK and $T_{i}=0.05$ K. All the parameters are scaled with $\omega_{a}$ in the numerical results.}
  \label{fig:fig5}
\end{figure}
 Our results suggest that bath spectrum filtering can provide an alternative to typical reservoir engineering schemes to create non-classical states. Bath spectrum filtering does not require external control for the desirable engineering of the
dissipative dynamics. Hence, it can be a promising route for creating exotic quantum states, useful for metrology or quantum computation~\cite{Ac_n_2018}.
\\
\acknowledgements
We thank Andr\'e Xuereb for the fruitful discussions. 

\appendix

\section{Simultaneous entanglement and cooling of the resonators}\label{App:A}

In a typical optomechanical system, simultaneous entanglement and cooling of the resonators are not possible. The reason is that cooling of the resonators requires beam splitter interaction, and entanglement generation between the resonators is possible with two-mode squeezing interaction. Depending on the input laser drive frequency choice, one of these two interactions can be made resonant. Accordingly, entanglement and cooling occur in two separate sets of system parameters~\cite{RevModPhys.86.1391}. On the contrary, in our scheme, both entanglement and cooling of the resonators can co-occur. This is one of the key features of our scheme, and to the best of our knowledge, it is not reported in the bosonic out-of-equilibrium entanglement generation schemes too. 

To investigate the entanglement and cooling of the resonators, after tracing the ancilla system A, Eq.~(\ref{eq:filtMEM}) is given by

\begin{eqnarray}
\frac{d\tilde{\rho}}{dt} &=& \sum_{i}({\gamma}_{i,\downarrow}D[\tilde{b}_{i}] + {\gamma}_{i, \uparrow}D[\tilde{b}_{i}^{\dagger}]+{\gamma}_{d}D[\tilde{b}^{\dagger}_{i}\tilde{b}_{i}])  \nonumber \\ &+& {\Gamma}_{\downarrow}D[\tilde{b}_{1}\tilde{b}_{2}] + {\Gamma}_{\uparrow}D[\tilde{b}^{\dagger}_{1}\tilde{b}^{\dagger}_{2}],
\end{eqnarray} 
here, we define
\begin{eqnarray}
{\gamma}_{i, \downarrow} &=& {f}_{i}(\omega_{i}), \qquad  {\gamma}_{i, \uparrow} = {f}_{i}(-\omega_{i}), \nonumber \\
{\Gamma}_{\downarrow} &=& \alpha^3 \tilde{f}(-\omega_{-})\langle{\tilde{n}}_{a}+1\rangle, \nonumber \\
{\Gamma}_{\uparrow} &=& \alpha^3\tilde{f}(\omega_{-})\langle\tilde{n}_{a}\rangle, \nonumber \\
{\gamma}_{d} &=& \alpha^3(\tilde{f}(\omega_{a})\langle\tilde{n}_{a}\rangle + \tilde{f}(-\omega_{a})\langle{\tilde{n}}_{a}+1\rangle).
\end{eqnarray} 
For sufficiently low excitations $\bar{n}_{i}$ in the resonators and $T_{h}\gg T_{i}>T_{c}$, the simultaneous cooling rate $\Gamma_{\downarrow}$ becomes larger than the simultaneous $\Gamma_{\uparrow}$ and the individual heating rates $\gamma_{i, \uparrow}$ of the resonators~\cite{Naseem2021}. Accordingly, the cooling of the mechanical resonators is accompanied by entanglement generation (see Sec.~\ref{subsec:NDR}). This analysis is confirmed by the numerical results presented in Fig.~\ref{fig:fig5}, which show that the cooling and entanglement co-occur in our scheme. Due to the different frequencies of the resonators, cooling rates are different~\cite{Genes_2008, Naseem2021}. 
 For the optimal performance of our scheme, the resonators should be considered in their ground states. However, this is not a requirement, as
  Fig.~\ref{fig:fig5} shows that the thermal state with sufficiently low excitations $\bar{n}_{i}$ can also lead to entanglement between the resonators.

\bibliography{ResEntg.bib}

\end{document}